\documentstyle[epsf,prl,aps]{revtex}

\newcommand{\model} {$u(\vec{x},t)$}
\newcommand{\data}  {$v(\vec{x},t)$}
\newcommand{\nphi}{K} 
\newcommand{\ncom}{N} 
\begin{document}
\title{The identification of continuous, spatiotemporal systems}
\author{H. Voss$\dagger$, M. J. B\"unner$\ddagger$, and  M. Abel$\dagger$}
\thanks{published in: Phys. Rev. E {\bf 57} (1998) 2820}
\address{$\dagger$ Institut f\"ur Theoretische Physik, Universit\"at Potsdam,
          Am Neuen Palais 10, D-14415, Potsdam, Germany.\\
         $\ddagger$Max--Planck Institut f\"ur Physik komplexer Systeme,
        N\"othnitzer Str. 38, D--01187 Dresden, Germany.}
\maketitle
\begin{abstract}  
We present a method for the identification of continuous, spatiotemporal  
dynamics from experimental data.
We use a model in the form of a partial differential equation and
formulate an optimization problem for its estimation from data.
The solution is found  as a multivariate nonlinear regression
problem using the ACE--algorithm.
The procedure is  successfully applied to data, obtained by simulation
of the Swift--Hohenberg equation. 
There are no restrictions on the dimensionality of  the investigated
system,
allowing for the analysis of high-dimensional chaotic as well
as transient dynamics.
The demands on the experimental data are discussed as well
as the sensitivity of the method towards noise.
\end{abstract} 
\pacs{\hspace{1.9cm}  PACS numbers: 07.05.Kf, 05.45.b}
 

The unstable dynamics observed in  spatially extended systems
attracted huge experimental and theoretical research activity
in the last decades (see~\cite{Cross93,Manneville90} and references
therein).
Progress has been achieved by describing the dynamics in the vicinity of 
bifurcations with the help of universal amplitude equations,  
vastly reducing the complexity of the involved models. 
Additionally, the research has concentrated on the classification of the 
observed instabilities  and the resulting patterns, and the
investigation of  scaling laws and intermittency effects~\cite{Cross93}.
For most investigations the models for spatiotemporal systems
arise from mainly theoretical considerations and  their validity is
affirmed by the comparison with experimental measurements.
Here we address the problem of finding a model, which describes the dynamics of
an observed system, directly from experimental data.
For systems which exhibit temporal low--dimensional chaotic motion this was
accomplished with the help of nonlinear maps~\cite{Abarbanel93}.
Other general methods rely on some sort of mode--expansion and were
successfully applied~\cite{Uhl-Armbruster-Berkooz}.
A different approach consists in a nonparametric model identification
as proposed for systems with a time delay feedback~\cite{PRE96a,kurths-voss-97}.

In this letter we extend that approach to
the identification of the underlying evolution equation of
spatially extended systems. 
At first we formulate an optimization problem
for finding a  model equation from the data. Then, we rewrite the
equations in the form of a multivariate nonlinear regression problem. 
As a last step, we use a novel kind of numerical algorithm for solving
the problem.
Our approach  does not
include any parameter dependencies, rather those are delivered as a
by-product. 
We  discuss the identification of homogeneous and autonomous 
partial differential equations, but
 emphasize  that
the ideas are quite general and can be applied to other problems like
finite-difference equations, coupled-map lattices or
integro--differential equations.

We assume the dynamics of the system under consideration to be
governed by a PDE of the form
\begin{equation} 
\label{pde} 
{\cal F}[\vec{\bigtriangledown},\partial_t,\vec{u}(\vec{x},t)]=0, 
\end{equation} 
where $\vec{u}$ is the field variable with $N$ components, $ \cal F$ is an
$N$--dimensional function of $\vec{u}(\vec{x},t)$ and its 
spatial and temporal derivatives, with $t$ and $\vec{x}$
the temporal and spatial variables, respectively.

To ease the analysis and the representation,
we only consider systems with a single component
$u$  and at most two spatial dimensions.
We note that the following considerations are valid in principle also
for multi--component systems and higher dimensions in space.

In the following, we discuss the procedure to estimate a PDE of the
form (\ref{pde}) from experimental data.
Therefore, we distinguish between the solution  \model\ and the
data \data.
For the sake of simplicity, we denote both the continuous space--time
variables of the model field and the discrete space--time variables of
the data by the same symbols $\vec{x}$ and $t$.
Considering the data $v$ as  random variables, we
act on a probability space and denote all entities estimated from the data by a hat
$\,\hat{\cdot}\/$\ .

Since  one can get only an estimate
for the true function $\cal F$ from the data $v$,
all one can achieve is to estimate the analogue of Eq.~(\ref{pde})
\begin{equation}\label{pde-est}
\hat{\cal F}[\hat{\bigtriangledown},\hat{\partial_t},
{v}(\vec{x},t)]=0\;,
\end{equation}
where $\hat{\cal F}$ is the estimate of $\cal{F}$ and
the derivatives have been substituted by estimates computed from the  data~$v$.

To obtain $\hat{\cal F}$, we formulate from
(\ref{pde})  the following optimization problem~\cite{Zeidler-85}:
\begin{equation}\label{norm1}
\inf_{{\cal F}} \|{\cal F} \| = e^2.
\end{equation}
The optimization lies in varying ${\cal F}$ until $e^2$
converges to the infimum.

The function $\cal F$ is defined as a function of operators,
e.g. $\nabla^2,\, \partial_t,\, {\rm id}, \ldots$.
We denote the set of constituting operators by
$\{{\cal O}_i\}_{i=1,\ldots,\nphi}$.
Note that in our definition of the differential operators ${\cal O}_i$
also any product terms like $u^2\partial_{xx}, \partial_t\partial_x$
are included. 
Since we consider a nonlinear problem, it is useful to split the function
$\cal F$ into sub--functions $\Phi_i$, which have
 as arguments ${\cal O}_i u$,
 e.g.~$\Phi(\partial_{xx}u)=(\partial_{xx}u)^\alpha$ or $\Phi({\rm
 id}u) = u^\beta$.
These $\Phi_i$ are elements of some class  of functions
$S$, which has to be specified according to the problem.
Thus, the final representation of Eq. (\ref{pde}) reads
\begin{equation}\label{sum-phi}
{\cal F} = \sum_{i=0}^\nphi \Phi_i({\cal O}_i u)=0.
\end{equation}
We  want to determine the constituents
${\cal O}_i u$ and  the functions $\Phi_i$ of $\cal F$ from a data
set~$v$. 
Therefore, we estimate at first ${\cal O}_i u$ from the data by finite
differencing or alternative schemes. Especially in the presence of
noise, Fourier methods or kernel estimation could be helpful.
The result consists of  $\nphi$  random variables
$\hat{\cal O}_i \equiv\widehat{{\cal O}_i u}$.
If the underlying PDE
is linear, i.e.  the function $\cal F$ a multilinear one, we could
solve the problem with linear regression methods.
Secondly, to obtain a solution
for the nonlinear problem, we solve
\begin{equation}\label{norm}
\inf_{\Phi \in S } \| \sum_{i=1}^\nphi \Phi_i(\hat{{\cal O}}_i) \| = e^2
\end{equation}
in varying the functions $\Phi_i$.
The result are the estimates  $\hat{\Phi}_i$.
Up to now, we anticipated to know the number and type of operators
${\cal O}_i$. For unknown systems, which are the ones we 
treat, it would be necessary to extend the number of operators $\nphi$ to infinity.
In practice one has to select a finite number.
Redundant terms  then deliver $\hat{\Phi}_i\approx0$ as result. 
It is important to find a
reasonable initial guess for the operators ${\cal O}_i$ of
Eq.~(\ref{sum-phi}). If there exists already some description for the
system in a special state,  
one starts with the operators which appear in
the known equations and tries to determine
additional terms which may appear when leaving that state.
This situation appears e.g. near some critical points, where one can
start with  already derived amplitude  equations.

In the case of data analysis, the optimization problem
(\ref{norm}) becomes a
multiple nonlinear regression problem which can be solved
using the {\em alternating conditional expectation algorithm\/}
(ACE)~\cite{Breiman85}.
It has already been successfully applied in related fields of
data analysis ~\cite{kurths-voss-97,Veaux89}.
In this case the norm of Eq.~(\ref{norm}) is the $L_2$ norm.
Using this algorithm, the functions $\Phi_i$ of Eq.~(\ref{sum-phi})
are seen as so-called {\em optimal transformations}~\cite{Breiman85},
which are defined to solve regression problems of the form
\begin{equation}
\label{multiregress}
E [(\Phi_0(\hat{\cal O}_0)-\sum_{i=1}^\nphi\Phi_i(\hat{\cal O}_i))^2]
\stackrel{!}{=}{\rm min}\;.
\end{equation}
The $\Phi_i$ are varied in the space $S$ of the Borel--measurable
functions
with the additional requirement of zero expectation, 
$E[\Phi_i]=0\;(i=0,\ldots,\nphi)$, and $E[\Phi_0^2]=1$.
The convergence of the ACE algorithm in the case of discrete samples
rather than random variables has also been proved in~\cite{Breiman85}.
The minimization of the expectation value~(\ref{multiregress}) is equivalent
to the calculation of the {\em maximal correlation} $\Psi$~\cite{footnote2}.
Instead of $e^2$ we use the maximal correlation as
a measure for the quality of the result:
$\Psi \in [0,1]$  equals 1
for perfect estimation, the smaller it is the worse is the estimation. 
The ACE--algorithm achieves the maximization by iteratively
transforming each $\hat{\cal O}_i$
by suitable, generally {\em nonlinear\/}, transformations
such as to yield a {\em linear\/} relationship
between the new random variables $\Phi_i(\hat{\cal O}_i)\/$.

As a first example,
we analyze data \data\ from the numerical integration of the
Swift--Hohenberg equation~\cite{Swift77}.
The model is of the form
\begin{eqnarray}\label{sheq1a}
\partial_tu &=& \left[ r-(\nabla^2 + k^2)^2 \right]u-u^3 \nonumber \\
            &=& (r-k^4)u-u^3-2k^2(\partial_{xx}+\partial_{yy})u \nonumber \\
            &&{}-(\partial_{xxxx}+\partial_{yyyy}+2\partial_{xxyy})u\;.
\label{sheq1b}
\end{eqnarray}
The global dynamics of the model can be derived from a
potential, such that the asymptotic time dependence is
trivial~\cite{Manneville90}.
Therefore, we analyze a transient state to have a sufficient
variation in the time derivative.
For the identification procedure we use data produced by
an explicit Euler integration scheme with a time step of $10^{-4}$,
a spatial discretization of $\Delta x=\Delta y=0.25$, and periodic
boundary conditions.
As initial conditions we choose uniformly distributed
independent random numbers from the interval $[-10,10]$.
The parameters are $r=0.1$ and $k=1$.

\begin{figure}[h]
\begin{center}
\epsfxsize=.53\textwidth
\leavevmode
\epsffile{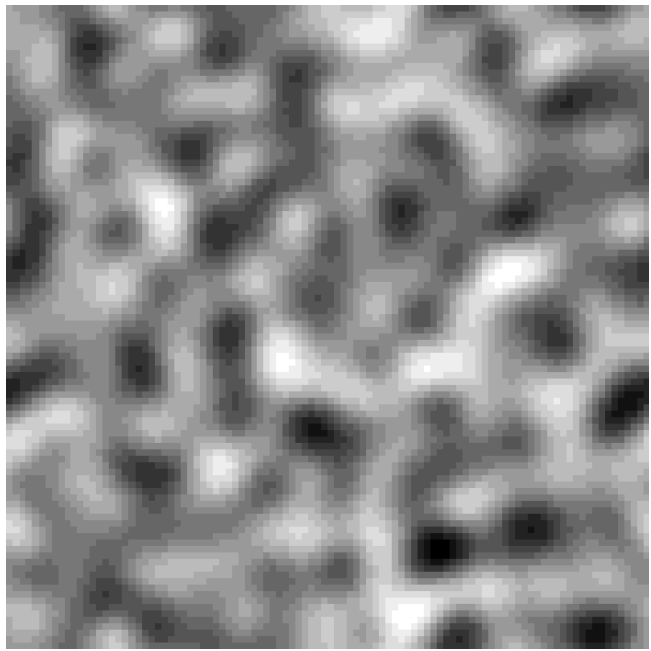}
\vspace{.3cm}
\caption[]{\label{shdata}
The data sample $v(x,y,t_0)$ for the central time point $t_0$
encoded in grey values (small values dark).
The horizontal and vertical axes are $x$ and $y$, respectively.}
\end{center}
\end{figure}

The differential operators are estimated by symmetric differencing
schemes,
e.g.
$\hat{\partial_t}v(\vec{x},t)=
(v(\vec{x},t+\Delta t)-v(\vec{x},t-\Delta t))/2\Delta t$.
Thus, to estimate the time
derivatives of first order in each spatial data point,
we need three consecutive "pictures" of data.
The  field size is $100\times 100$ points, i.e., the data set \data\
contains $3*10^4$ values.
The data for the central time point are shown in Fig.~\ref{shdata}.
In the following we can drop without ambiguity 
the hat~$\hat{.}\/$~.

To identify the unknown system, we
use an ansatz with non--mixed terms (like $\partial_x v$)
up to fourth order in the spatial derivatives.
To show how to handle mixed terms we additionally include the
terms $v\partial_xv$, $v\partial_yv$, $\partial_xv\partial_yv$.
\begin{eqnarray}\label{sheq2}
\Phi_0(\partial_tv)&=&
\Phi_1(v)+\Phi_2(\partial_xv)+\Phi_3(\partial_yv)\nonumber\\
&&+\Phi_4(\partial_{xx}v)+\Phi_5(\partial_{xy}v)+\Phi_6(\partial_{yy}v)
\nonumber\\
&&+\Phi_7(\partial_{xxx}v)+\ldots+\Phi_{10}(\partial_{yyy}v)\\
&&+\Phi_{11}(\partial_{xxxx}v)+\ldots
+\Phi_{15}(\partial_{yyyy}v)\nonumber\\
&&+\Phi_{16}(v\partial_xv)+\Phi_{17}(v\partial_yv)\nonumber\\
&&+\Phi_{18}(\partial_xv\partial_yv)\nonumber\;
\end{eqnarray}
with the nine redundant terms
$\Phi_2(\partial_xv)$, $\Phi_3(\partial_yv)$, $\Phi_5(\partial_{xy}v)$,
$\Phi_7(\partial_{xxx}v)$, $\Phi_8(\partial_{xxy}v)$,
$\Phi_9(\partial_{xyy}v)$, $\Phi_{10}(\partial_{yyy}v)$,
$\Phi_{12}(\partial_{xxxy}v)$, $\Phi_{14}(\partial_{xyyy}v)$,
$\Phi_{16}(v\partial_xv)$, $\Phi_{17}(v\partial_yv)$,
$\Phi_{18}(\partial_xv\partial_yv)$.
We choose this ansatz as a compromise between generality and
computational effort.
Comparing Eq.~(\ref{sheq1b}) with Eq.~(\ref{sheq2}), one expects
in particular the following for the solution of Eq.~(\ref{multiregress}):
Up to an arbitrary common factor,
$\Phi_0$ should be the identity,
$\Phi_1$ should be a polynomial of third order,
and for $i=4,6,11,13,15$ the $\Phi_i$ should be linear functions
$\alpha_i {\cal O}_i$. 
All other estimates should vanish.
Furthermore, we expect for the slopes of the linear functions, after
division by the slope of the l.h.s to remove the arbitrary common
factor, $\alpha_4=\alpha_6=\alpha_{13}=-2$, $\alpha_{11}=\alpha_{15}=-1$.

\begin{figure}
\begin{center}
\epsfxsize=.5\textwidth
\leavevmode
\epsffile{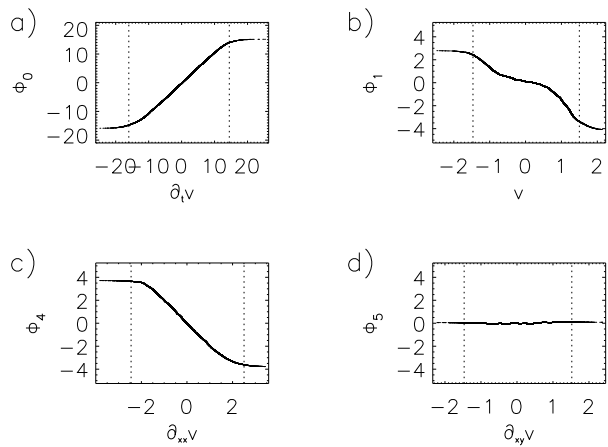}
\caption[]{\label{sh1}
Four exemplary estimates for the optimal transformations
$\Phi_0$, $\Phi_4$, $\Phi_5$, $\Phi_{16}$ are shown.
The linear occurrence of $\partial_tu$ and $\partial_{xx}u$ is
clearly recovered by $\Phi_0$ and $\Phi_4$, respectively.
The redundant terms $\partial_{xy}u$ and $u\partial_xu$
vanish approximately. 
The dotted lines mark the interval on the abscissa in which 98\% of
the data points are located.
Due to a very non--homogeneous distribution of the data
the optimal transformations outside the marked interval cannot  be
estimated reliably.
The results for the optimal transformations not shown here deliver
estimates of equal quality.
}
\end{center}
\end{figure}

\begin{figure}
\begin{center}
\leavevmode
\epsffile{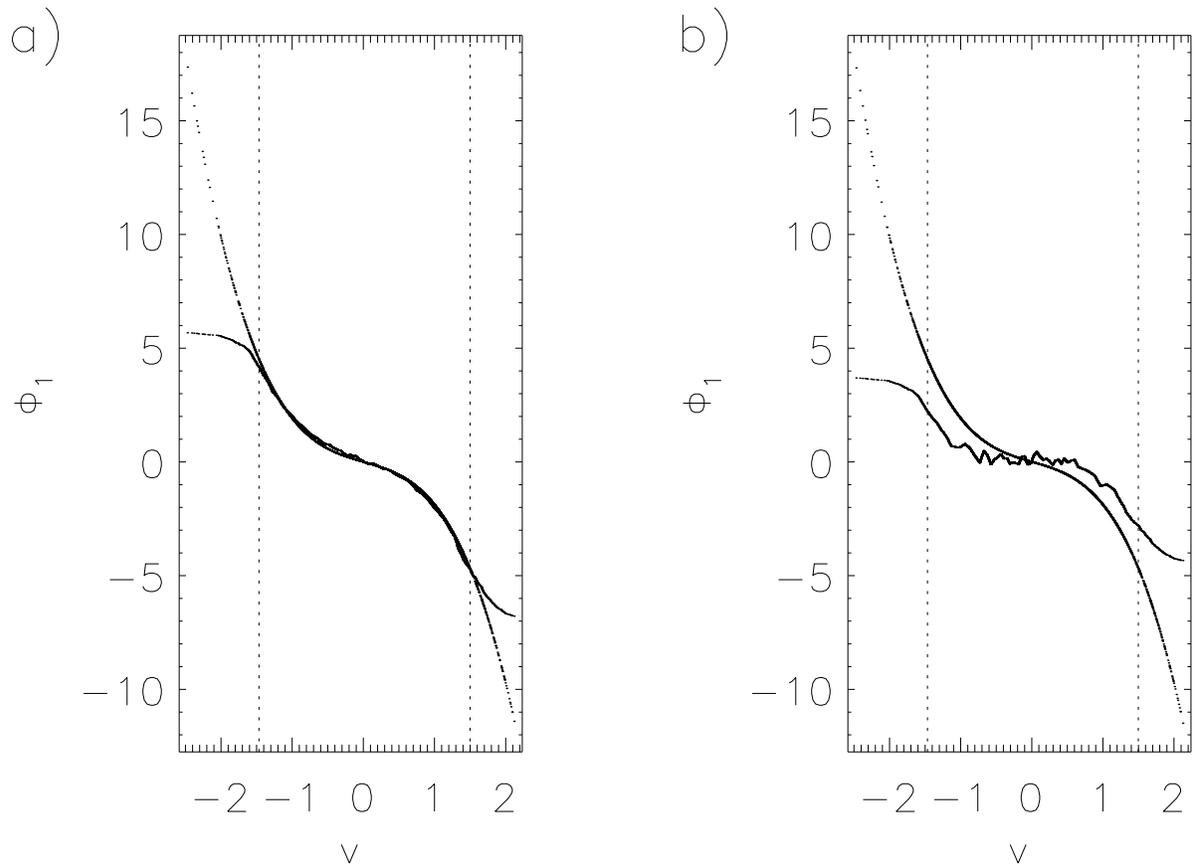}
\caption[]{\label{sh2}
The nonlinear term  of the Swift--Hohenberg equation~(\ref{sheq1a}).
a) The estimated and the exact nonlinearity $-0.9u-u^3$.
b) The same with noise, as explained in the text.
}
\end{center}
\end{figure}

Performing the ACE--algorithm, we find a maximal correlation of
$0.9993$
and optimal transformations as shown in~Fig.~\ref{sh1}.
All functions approximate the expected shape well, and the
terms which were expected to vanish are indeed very small compared to
the other ones.
Note that this does not mean that the values for
the redundant terms are vanishing
themselves but that these are independent from all the
other terms involved. The comparison of the slope of the linear
functions yields the possibility to estimate parameters;
we obtain
$\alpha_4=\alpha_6=-1.9$, $\alpha_{13}=-2.0$, $\alpha_{11}=\alpha_{15}=-1.0$.
The slopes of the terms which are expected to vanish are all smaller
than $0.03$ in absolute value.
Since it is difficult to estimate the errors of these values, it is
recommended to check the range of validity of the reconstructed model
by integration and comparison with the  dynamics.
In Fig.~\ref{sh2}  the nonlinearity $-0.9u-u^3$
of Eq.~(\ref{sheq1b}) is compared explicitly with the estimate, taking
now an ansatz with only the non--vanishing terms of the above result.
Inside a range of 98\%
of the data values this term can be
estimated with high accuracy.

>From a practical point of view,
it is essential to discuss the stability of the identification method 
against noise.
The effect of noise is to increase the errors of the estimates for the
partial differential operators applied to the data, and therefore also
the value of the error estimate  
$e^2$.
The identification detects the remaining correlations of  
the vector field and its derivatives via the function ${\cal F}$.
In general, if the vector field and its 
derivatives are still strongly correlated for a reasonable low 
noise level of a noisy system,
the minimum according to  Eq. 
(\ref{norm}) should still be detectable.

To examine the stability against noise,
we disturb each data point \data\
by additive Gaussian white noise with a
standard deviation of 0.5\%
of the data standard deviation.
The estimates of the partial derivatives are 
disturbed severely due to error propagation.
In spite of this, using again Eq.~(\ref{sheq2}),
the overall shape of all functions can still be recovered satisfactory, while being
distorted (Fig.~\ref{sh2}).
The maximal correlation has decreased to 0.926.

Two other systems which were analyzed with
similar success as the example above, are:
1. The Kuramoto--Sivashinsky equation
(in the form as given in Ref.~\cite{Manneville90})
in the fully chaotic regime.
2. A reaction--diffusion  system with two components~\cite{Baer93},
where the nonlinearity in the inhibitor dynamics is
a non--analytic function.
We reconstructed that function with high accuracy for different states
like a single rotating spiral and spiral--defect chaos.

Summarizing our experiences, we find the following requirements on the
data:
1) In order to identify a system with \ncom\ components it is sufficient
to measure \ncom\ independent variables.
2) The sampling of the vector fields in space as well as in time has to be  
appropriate to estimate the partial derivatives with respect to time
and space properly.
3)
In the preceding example we analyzed data at one time point but for an
extended region in space.
In other situations it may be useful to perform the analysis at a
single spatial point but for a longer interval in time, or even a
combination of both procedures.
However, as the example shows, even a small amount of data can be
sufficient.
4) Since the reconstructed functions can only be determined at
points which are attained by the data and the estimates for the
differential operators, the data have to show a sufficient variability
in space and time.
5) The inference from the data to a PDE is often not unique.
But within the errors of the algorithm, our method reasonably
reconstructs the dynamics on the trajectory given by the data.
6) We believe that due to the statistical nature of the method
a higher  noise level can typically be compensated by a larger data set.

We would like to stress the fact that the procedure presented above is
essentially independent  of the part of phase space the system moves on
and the boundary and  initial conditions.
Thus we do not  require the dynamics to be close to or away from  
an attractor.
This kind of  analysis is thus
applicable in the case of high-dimensional chaotic motion, as
exhibited by nonlinear spatially extended systems, as well as
transient motion.   
We did not discuss non-homogeneous or non-autonomous evolution laws,
but in principle the arguments apply equally well, with some minor
modifications, for those systems.

In conclusion, we presented a method for the identification of spatiotemporal systems
by numerical reconstruction of the PDE which describes the system.
The identification procedure consists in solving an optimization
problem which results in a nonlinear regression problem.
Using the ACE--algorithm we showed that this  task can
indeed be solved in the case of numerical examples.
We consider  the method as a useful tool for the analysis of
spatiotemporal systems and expect it to find a large variety of
applications.
The limits of this method and the requirements on the data have been
discussed. 
Future work will concentrate on an extension of the ACE--algorithm to
the solution of multi--component problems,
and on the application to real data.
Extensions to more general systems are envisaged.

We acknowledge helpful discussions with H.\ Kantz, J.\ Kurths, J.\ Parisi,
A.\ Pikovsky, M.\ Zaks, and financial support by the
Max--Planck--Gesellschaft.

\vspace{1.5cm}

\end{document}